We derive and discuss expressions for the temperature-dependent electrostatic polarizabilities of off-center ions holding good under various experimental conditions. At low temperatures and field strengths, all of them reasonably reduce to values characteristic of phonon-coupled two-level systems. Prospects for further studies of the dispersive coupling are also considered.

## 1. Rationale

In a series of recent preprints, we considered the off-center behavior of small size $Li^+$ impurities in fcc alkali halide crystals [1-5]. Among the features discussed were the optical properties, the 3D and 2D Hamiltonians, the exact Mathieu eigenfunctions in 2D, and the reorientational rate of 2D rotors. The analytic results were compared with calculations by the extended Hückel method.

Throughout these papers, we used a second-order perturbation definition of the polarizability which is only correct at small dipolar parameters. Now we want to check just how far the definition could be extended so as to incorporate higher values. Of particular importance are the temperature-dependent polarizabilities, since these should control the processes running above zero point. We will also be particularly interested in high polarizabilities beyond second-order giving rise to dispersive binding energies of considerable strength, as proposed elsewhere [6].

## 2. General definitions

The splitting of a normal lattice site into a domain of off-center sites, to be occupied by a small-radius impurity ion, leads to the occurrence of an off-site ellipsoid. Indeed, on minimizing the adiabatic energy

$$E_L(\{Q_i\}) = (1/2)\{\sum_i K_i Q_i^2 - [\sum_i (2G_i Q_i)^2 + E_{\alpha\beta}^2]^{1/2}\} \qquad (1)$$

with respect to $Q_i$ we arrive at the equation

$$Q_x^2/Q_{x0}^2 + Q_y^2/Q_{y0}^2 + Q_z^2/Q_{z0}^2 = 1 \qquad (2)$$

where the semiaxes read

$$Q_{x0} = [(2E_{JTx}/K_x)(1-\eta_x^2)]^{1/2}, \quad \eta_x = E_{\alpha\beta}/4E_{JTx}$$

$$Q_{y0} = [(2E_{JTy}/K_y)(1-\eta_y^2)]^{1/2}, \quad \eta_y = E_{\alpha\beta}/4E_{JTy}$$

$$Q_{z0} = [(2E_{JTz}/K_z)(1-\eta_z^2)]^{1/2}, \quad \eta_z = E_{\alpha\beta}/4E_{JTz} \qquad (3)$$

This ellipsoid is the locus of metastable spatial positions of the impurity ion which are displaced off-center with respect of the normal lattice site. An off-center instability clearly occurs about the normal lattice site at $\eta_x, \eta_y, \eta_z < 1$ where $E_{JTi} = G_i^2/2K_i$ are Jahn-Teller energies, $G_i$ are the electron-vibrational mode coupling constants, $K_i = M\omega_i^2$ are the stiffness constants, and $\omega_i$ are the bare vibrational frequencies associated with $Q_i$.

The off-center ellipsoid is polarizable electrostatically. Its coupling energy with an external electric field is obtained from

$$U_c = \tfrac{1}{2} \alpha\, F.F = \tfrac{1}{2} \sum_i \alpha_{ij} F_i F_j \qquad (4)$$

where $\alpha$ is the polarizability tensor of the off-center ellipsoid and F is the electric field strength.

The off-site ion traverses across the off-center volume and around the normal lattice site which modifies its (ionic) polarizability. Accordingly, $\alpha$ is the polarizability tensor of the off-center ellipsoid rather than the one of an isolated ion.

### 3. Energy of a two-level system

Assuming Boltzmann statistics, the energy of a two-level system in an external field F is easily found to be

$$E(T,F) = -\delta_F \tanh(\delta_F/k_B T) \qquad (5)$$

where $\delta_F \equiv \delta(F) = \sqrt{(\delta_0)^2 + (\mathbf{p}_E.\mathbf{F})^2}$ is the field-dependent tunneling splitting, while $\delta_0 \equiv \delta(0)$ is the splitting in the absence of a field. Inserting into (5) we get

$$E(T,F) = -\delta(F) \tanh(\delta(F)/k_B T) = -\sqrt{(\delta_0)^2 + (\mathbf{p}_E.\mathbf{F})^2}\, \tanh(\sqrt{(\delta_0)^2 + (\mathbf{p}_E.\mathbf{F})^2}/k_B T) \qquad (6)$$

At weak fields, $\mathbf{p}_E.\mathbf{F} \ll \delta_0$, $E(T,F) \sim -\delta_0[1 + \tfrac{1}{2}(\mathbf{p}_E.\mathbf{F}/\delta_0)^2]\tanh(\delta_F/k_B T)$ and we obtain from (5):

$$E(T,F) \sim -\delta_0 \tanh(\delta_0/k_B T) - \tfrac{1}{2}[(\mathbf{p}_E.\mathbf{F})^2/\delta_0]\tanh(\delta_F/k_B T)$$

$$\cong -\delta_0 \tanh(\delta_0/k_B T) - \tfrac{1}{2}[(p_E^2/\delta_0)\tanh(\delta_0/k_B T)]\,F^2 \qquad (7)$$

Here the quantity

$$\alpha(T) = \tfrac{1}{2}(p_E^2/\delta_0)\tanh(\delta_0/k_B T) = \alpha(0)\tanh(\delta_0/k_B T) \qquad (8)$$

is the low-field temperature-dependent electrostatic polarizability of a two-level system where $\alpha(0) = \tfrac{1}{2}(p_E^2/\delta_0)$ is the low-field polarizability at zero point.

At high fields $\mathbf{p}_E.\mathbf{F} \gg \delta_0$, the relevant equations are nonlinear. From eqn.(6) we find instead

$$\alpha(F) = \sqrt{(\delta_0)^2 + (\mathbf{p}_E.\mathbf{F})^2}\, \tanh(\sqrt{(\delta_0)^2 + (\mathbf{p}_E.\mathbf{F})^2}/k_B T) \sim p_E^2/k_B T \qquad (9)$$

which is the reorientational polarizability of particles with permanent dipole moments [7].

### 4. Examples for temperature dependent off-center polarizabilities

To derive the components of the polarizability tensor, we should specify its coupling to the $Q_i$ modes. For this purpose we consider a few examples of off-center configurations [7].

### 4.1. Two-site flip-flops – a configurational segment in 1D

A familiar example is the two off-site flip-flops of the ammonia molecule $NH_3$ [7]. The central higher energy configuration places all the four atoms on the vertexes of two equilateral triangles with a common H-H basis, the remaining N and H then occupy the two free vertexes facing the basis. However, this configuration being unstable, the higher symmetry is broken by displacing the N atom further away from the common basis, the residual non-deformed triangle being formed by the three remaining H atoms. As a result, two equivalent lower energy substitutes occur as left-hand and right-hand ones. Transitions from left to right and back correspond to 1D flip-flops. From eqn. (8), the off-center polarizability of the two-site system is:

$$\alpha(T) = (p_E^2/2\delta_0) \tanh(\delta_0 / k_B T) \tag{10}$$

Here $p_E$ is the electron states' mixing dipole, $\delta_0$ is the vibronic tunnelling splitting whose finite value makes the left-to-right interwell transitions possible at all.

### 4.2. Four off-center sites in-plane – a configurational circle in 2D

Coupling to the odd-parity bending modes $Q_x$ and $Q_y$ of the two <110> $Cl^-$–$Li^+$–$Cl^-$ in-plane bonds results in four off-center sites within the basal (x,y) plane of an $F_A$ center with [7]:

$$\alpha_{(x,y)} = [p_E^2(1 - \eta_E)/3\delta_0]\{(\delta_0/k_B T)\exp(-\delta_0/k_B T) + \sinh(\delta_0/k_B T)\}/$$
$$\{\exp(-\delta_0/k_B T) + \cosh(\delta_0/k_B T)\} \tag{11}$$

as defined by means of the mixing dipoles

$$p_E = p_{(x,y)} \delta n_{(x,y)} = <a_{1g}|ex(y)|t_{1ux(y)}> \delta n_{(x,y)}$$

while the tunnelling splitting $\delta_0$ associated with the transfer between off-center sites is defined by means of the respective JT energies, elastic and coupling constants [8]:

$$\delta_0 = 2E_{JTE}\, \eta_E\, (1 - \eta_E)/\exp(u_E^2)$$
$$u_E = \{(2E_{JTE}/\eta\omega_E)[1 - \eta_E^2]^{1/2}\}^{1/2}\,. \tag{12}$$

$\delta n_{x(y)} = \delta n_{t_{1u}\, x(y)} - n_{a_{1g}}$ are the differences in occupation numbers between $t_{1ux(y)}$ and $a_{1g}$. Hereafter we set $\eta \equiv h/2\pi$, while $\eta_E = E_{gap}/4E_{JT}$ is the two-level gap parameter.

$\alpha_{(x,y)}$ is the electrostatic polarizability of the off-center $Li^+$ impurity in (x,y) plane immediately under the F center at [100] which leads to an off-center polarizable circle. It can be essential in experiments on the electrostatic behavior of $Li^+$ impurities and $F_A$ centers in alkali halides, e.g. in paraelectric resonance [9]. It displays a specific temperature dependence with a low-temperature branch at $k_B T \ll \delta_0$:

$$\alpha_{(x,y)}(0) = p_E^2(1 - \eta_E^2)/3\delta_0 \,, \tag{13}$$

as well as a high-temperature branch at $k_B T \gg \delta_0$ which gives an orientational polarizability [7]:

$$\alpha_{(x,y)}(T) = p_E^2 (1 - \eta_E^2)/k_B T, \tag{14}$$

Equation (11) has been derived earlier and can be obtained from a well-known textbook. We now derive an energy and consider the electrostatic energy equation in 2D (four off-center sites):

$$E(T,F) = -\delta(F)[3/2\, \sinh(3/2\, \delta/k_BT) + \tfrac{1}{2} \sinh(\tfrac{1}{2}\, \delta/k_BT)] / [\cosh(3/2\, \delta/k_BT) + \cosh(\tfrac{1}{2}\, \delta/k_BT)]$$

leading easily to

$$E(T,F) = -\tfrac{1}{2}\, \delta(F)\, \tanh x\, [5 + 6\sinh^2 x] / [2\cosh^2 x - 1], \quad x = \tfrac{1}{2}\, \delta/k_BT \quad (15)$$

From $\delta(F) = \sqrt{(\delta_0)^2 + (\mathbf{p}_E.\mathbf{F})^2} \sim \delta_0 [1 + \tfrac{1}{2} (p_E/\delta_0)^2 F^2]$ the four-site low-field polarizability is

$$\alpha(T) = (p_E^2/4\delta_0)\, \tanh(\tfrac{1}{2}\delta_0/k_BT)[5 + 6\sinh^2(\tfrac{1}{2}\delta_0/k_BT)] / [2\cosh^2(\tfrac{1}{2}\delta_0/k_BT) - 1] \quad (16)$$

At low field and zero point, equation (9) leads to the polarizability of a two-level system $\alpha(0) = \tfrac{1}{2}\, p_E^2/\delta$ taking account of $\delta \equiv \delta(F) = \sqrt{\delta(0)^2 + (\mathbf{p}_{12}.\mathbf{F})^2} \sim \delta(0) + \tfrac{1}{2} (\mathbf{p}_{12}.\mathbf{F})^2/\delta(0)$ which gives $x \sim \tfrac{1}{2} [\delta(0) + \tfrac{1}{2} (\mathbf{p}_{12}.\mathbf{F})^2/\delta(0)] / k_BT = \tfrac{1}{2} [\delta(0) + \alpha(0)\, F^2] / k_BT$ and $\tanh x \sim \{\tanh x_0 + \tfrac{1}{2}\, \alpha(0)\, F^2 / k_BT\}/ \{1 + \tanh x_0\, \tfrac{1}{2}\, \alpha(0)\, F^2 / k_BT\} \sim \tanh x_0 + \tfrac{1}{2}\, \alpha(0)\, F^2 / k_BT$ for $x_0 = \tfrac{1}{2}\, \delta(0)/k_BT$.

From eqn.(11) at $\delta_0 \gg k_BT$ we get $\alpha_{(x,y)} = [p_E^2(1 - \eta_E^2)/3\delta_0]\, \tanh(\delta_0/k_BT)$ which implies that at low temperatures the four off-center sites become identical to two 1D pairs polarizable by eqn. (10). This conclusion is understandable in view of the configurational physics and geometry involved.

Generally, the off-center polarizability is independent of the applied electric field at low fields. To estimate the effect of high fields, we remind that the tunneling splitting becomes dependent on the field as $\delta_F = \sqrt{\delta_0^2 + \mathbf{p}_E.\mathbf{F}^2}$ where $\mathbf{p}_E$ is the field coupling dipole. As $\mathbf{F}$ grows in magnitude, we get from eq. (5) $\alpha_{(x,y)}(T,F) \to \alpha_{(x,y)}(0,F) = p_E^2(1 - \eta_E^2)/3\delta_0$ for $F \to \infty$: the field drives the polarizability towards its zero-point value.

### 4.3. Six off-center sites - configurational sphere in 3D

This is the general case of a substitutional $Li^+$ impurity going off center over six <111> positions around a normal cationic site in an alkali halide lattice. An off-center sphere results from equation (3) for the isotropic crystalline medium. In the presence of an external electric field the sphere is polarizable and its energy in the field reads:

$$E(T,F) = -\delta(F)[\tfrac{1}{2} \sinh(\tfrac{1}{2}\, \delta/k_BT) + 3/2\, \sinh(3/2\, \delta/k_BT) + 5/2\, \sinh(5/2\, \delta/k_BT)] /$$

$$[\cosh(\tfrac{1}{2}\, \delta/k_BT) + \cosh(3/2\, \delta/k_BT) + \cosh(5/2\, \delta/k_BT)]$$

which, following certain manipulations, reduces to

$$E(T,F) = -\tfrac{1}{2}\delta(F)\tanh x\, [35 + 112\sinh^2 x + 80\sinh^4 x]/[3 - 16\cosh^2 x + 16\cosh^4 x], \quad x = \tfrac{1}{2}\delta/k_BT \quad (17)$$

The corresponding six-sites low-field electrostatic polarizability reads

$$\alpha(T) = (p_E^2/4\delta_0)\, \tanh(\tfrac{1}{2}\delta_0/k_BT)[35 + 112\sinh^2(\tfrac{1}{2}\delta_0/k_BT) + 80\sinh^4(\tfrac{1}{2}\delta_0/k_BT)] /$$

$$[3 - 16\cosh^2(\tfrac{1}{2}\delta_0/k_BT) + 16\cosh^4(\tfrac{1}{2}\delta_0/k_BT)] \quad (18)$$

At zero point and low field the vibronic energy tends to $E(0,F) = \delta\tanh(\delta/k_BT)$ leading to the polarizability $\alpha(0) = \frac{1}{2}[p_E^2/\delta(0)]$ of a two-level system. At not too low a temperature, however, $E(T,F) = \delta(F)\tanh^3(\delta/k_BT)$ yielding $\alpha(T) = \frac{1}{2}[p_E^2/\delta(0)]\tanh^3(\delta/k_BT)$.

## 5. Further developments

Throughout this study, we found it hard to develop a method for predicting any grand values of the vibronic polarizabilities apart from those inferred from the formal application of second-order perturbation to dipole moments and vibronic tunneling splittings in phonon-coupled two-level systems. However, enhanced polarizabilities have been inferred indirectly from exciton matter appearing in atmospheric discharges as well as in dispersive binding of molecular systems [6].

In pursuing the problem, we elaborated methods for deriving temperature-dependent electrostatic polarizabilities of 1D, 2D, and 3D systems off-centered with respect to the normal lattice sites. The results obtained might be useful for checking the conclusions drawn in earlier papers [1–5,9–12] based on the zero-point polarizabilities. We have now seen that accounting for the temperature effects tends to reduce the polarizabilities and therefore the dispersive binding [6]. However, any conclusion to the effect that dispersive binding may be generally weak would be premature as long as it is based on second-order perturbation. We remind that 2[nd] order appears as the lowest-order perturbation solution as long as the 1[st] order is vanishing in the vibronic energy equation. Even within the framework of perturbation theory, attempts have been made to take account of third- and higher-order terms in order to explain the higher polarizabilities deduced from experiment [13]. Parallel attempts have also been made to address contributions to the binding energy arising from quadrupole and higher-order multipole effects.

The frequency-dependent polarizabilities are another, though not yet well-appreciated, chapter of the dispersive interactions.

Other well known factors not accounted for presently are retardation and damping [13]. It is now well realized that the dispersive binding occurs by virtue of the coupling of induced dipoles on constituent atoms and, therefore, should be more effective at shorter separations so as to preserve phase coherence.

Another curious challenge for the future is the magnetic dispersive binding which requires deriving induced magnetic dipoles $\mu = \chi B$ and polarizabilities $\chi$ in a similar way as we have done so for electric dipoles $p = \alpha F$ and polarizabilities $\alpha$ [13].

Finally, the dispersive binding being universal, it can also arise in soft condensed media and even in the gaseous phase with a low energy gap. This could have some effect on the primary binding of the elements at the early stages of the Universe. As solid state media are formed, however, the pressure effects tending to increase the energy gap, the dispersive binding is losing ground and other types of a chemical bond gain relative strength.

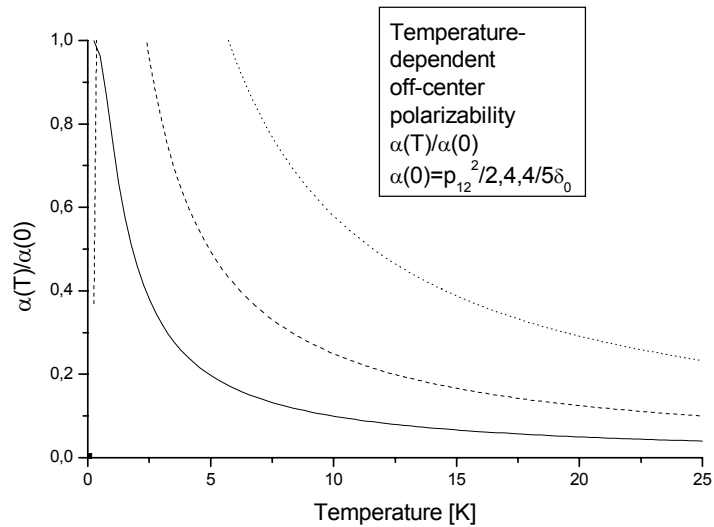

Figure 1: Temperature-dependent off-center polarizabilities in 1D (solid), 2D (dashed), and 3D (dotted). These dependencies followed from equations (10), (16) and (18), respectively. At lower temperatures, the polarizabilities exhibit their zero-point values $\alpha(0)$, at higher temperatures they tend to the orientational polarizabilities $\propto 1/T$.